# Discovering Patterns in Biological Sequences by Optimal Segmentation

**Joseph Bockhorst**[*]
Electrical Engr. and Computer Science
University of Wisconsin - Milwaukee
Milwaukee, WI 53211

**Nebojsa Jojic**
Microsoft Research
Redmond, WA 98052

## Abstract

Computational methods for discovering patterns of local correlations in sequences are important in computational biology. Here we show how to determine the optimal partitioning of aligned sequences into non-overlapping segments such that positions in the same segment are strongly correlated while positions in different segments are not. Our approach involves discovering the hidden variables of a Bayesian network that interact with observed sequences so as to form a set of independent mixture models. We introduce a dynamic program to efficiently discover the optimal segmentation, or equivalently the optimal set of hidden variables. We evaluate our approach on two computational biology tasks. One task is related to the design of vaccines against polymorphic pathogens and the other task involves analysis of *single nucleotide polymorphisms* (SNPs) in human DNA. We show how common tasks in these problems naturally correspond to inference procedures in the learned models. Error rates of our learned models for the prediction of missing SNPs are up to 1/3 less than the error rates of a state-of-the-art SNP prediction method. Source code is available at www.uwm.edu/∼joebock/segmentation.

## 1 INTRODUCTION

As the amount and kinds of spatial and temporal sequence data increases, so will the importance of computational methods for discovering patterns in sequences. An important problem in computational biology is to discover correlations among nearby positions in biological sequences. The immune system, for example, recognizes short stretches of amino acids in pathogen proteins. Thus understanding patterns of sequence diversity on this scale is an important problem in vaccine design. In this paper we describe how to learn the optimal segmentation of a sequential data set into non-overlapping segments. Our approach involves learning a probabilistic model over sequences.

One popular and powerful approach in sequence processing is to learn a probabilistic model, often a hidden Markov model (Rabiner, 1989) or related Bayesian network (BN), in which the observed variables represent sequences. An important aspect of these techniques is that they capture dependencies among sequence elements using hidden variables (HVs), variables whose values are infrequently or never observed. Hidden variables simplify otherwise apparently complex dependencies among the observed variables. This simplification has two key advantages: (i) a BN with correctly placed HVs usually has fewer parameters, which can be more accurately estimated, than does a related BN without HVs that encodes the same conditional independencies, and (ii) HVs often lend an interpretation that assists understanding the domain. HVs are also useful in prediction tasks by, for example, labeling sequence positions. Our segmentation procedure discovers an optimal set of hidden variables.

Related work for discovering hidden variables includes the approach of Elidan et al. (2001) to discover unconstrained hidden variables. Their method first learns the BN structure over the observed variables using standard methods, and then considers candidate HVs to untangle "semi-cliques" of the learned network. The algorithm of Zhang et al. (2004) learns tree structured BNs that generalize naive Bayes models in which discovered HVs are restricted to internal nodes below the root class node and above the observed leaf nodes. Despite these constraints, the complexity of the search space prevents exhaustive search, and a greedy heuristic search is used.

---

[*] JB performed a portion of this work at Microsoft Research.

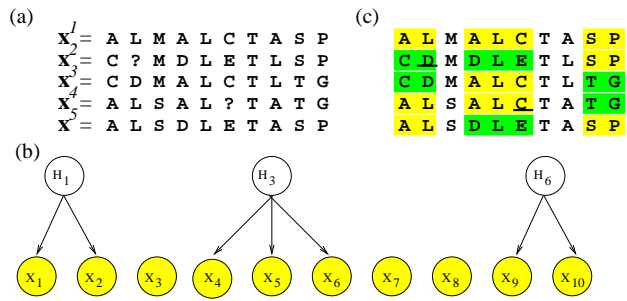

Figure 1: An example of our approach on a toy data set (a) with five length 10 sequences (amino acids in this case) that has two missing observations (the ?'s). We show in (b) the discovered hidden variables and corresponding BN structure. We require that each hidden (unshaded) variable connects to a contiguous *segment* of the observed (shaded) variables, and that no two segments overlap. Thus, the learned BN consists of a combination of independent mixture models, one per segment, and independent sequence positions (the unconnected observed variables). Using the learned model, we answer domain specific queries performing inference. In (c) the colors show the most likely types (i.e., HV values) for each segment, and, the predictions of the missing letters (underlined).

Unlike these approaches, which do not assert any *a priori* relationship among the observed variables, we restrict our attention to sequential domains in which the observed variables are ordered. We use this ordering to define a natural constraint on the HVs. Specifically, we only consider HVs that are parents of a contiguous segment of observed variables, and we do not allow the segments of any two HVs to overlap. Thus, each observed variable has either zero parents or a single hidden variable parent. Our approach finds the optimal number of HVs, their placement, and number of states (or cardinality) of each HV.

Figure 1 shows an overview of our approach. We take as input a set of $N$ aligned sequences $\mathbf{x}^1, ..., \mathbf{x}^N$ of discrete symbols from a finite alphabet. Each sequence $\mathbf{x}^j = (x_1^j, ..., x_L^j)$ has $L$ symbols, and $x_i^j$ refers to the $i^{th}$ symbol in sequence $\mathbf{x}^j$. We allow symbols in the input data set to be missing. Figure 1(a) shows example input sequences. We assume that the dependencies of interest are between nearby positions, and seek to explain these dependencies with HVs. Our approach scores each candidate HV position and cardinality (up to a maximum). We recover the maximum scoring set of non-overlapping HVs with a dynamic program. Figure 1(b) shows a likely BN learned from the input in Figure 1(a). Here, $X_i$ is the random variable for the symbol at position $i$, and the $H$'s are hidden variables. Given the learned model, we perform domain appropriate inference tasks. In Figure 1(c), for example, we predict missing symbols and assign types to the segments of each sequence.

The main contributions of this paper are:

- Our approach to discovering the optimal segmentation of a sequential data set. This involves finding an optimal set of hidden variables.

- An evaluation of three hidden variable scoring measures, MDL/BIC, Cheeseman-Stutz and cross-validation (CV). Our results indicate that the empirical method (CV) outperforms the other more theoretical approaches on the problem sets considered here. We elaborate on this point later.

- A description of how important domain specific queries naturally correspond to simple inferences in the learned model. On one key problem (predicting missing SNPs), our method yields a significant increase in accuracy over the state-of-the-art.

## 2 PROBLEM DOMAINS

Our work is motivated by problems arising in the analysis of biological sequences. In particular, this paper deals with the problem of discovering dependencies between nearby positions in a set of related sequences. Two applications of such methods are in the design of vaccines and analysis of single nucleotide polymorphisms.

### 2.1 VACCINE DESIGN

The original motivation for this work arose in the context of vaccine design. Despite considerable effort, safe and effective vaccines have not been developed for some of the world's deadliest diseases, including AIDS and malaria, which are each responsible for well over one million deaths annually (Lopez & Mathers, 2006). One key obstacle to the development of vaccines for these diseases stems from the *polymorphism* of the amino acid sequences of pathogen proteins recognized by the human immune system. Because these proteins can come in different forms, infections of the same disease caused by different organisms may have different immune signatures. Thus, an acquired immune response targeting one version of the pathogen may not be protective from others. For example if an immune response was learned based on the first two positions of the sequences in Figure 1(a), exposure to $\mathbf{x}^1$ would likely not induce protection against $\mathbf{x}^3$. Other recent approaches to vaccine design consider populations of pathogen sequences in an attempt to construct broadly effective vaccines (Jojic et al., 2005;

Nickle et al., 2003). These approaches require an understanding of the natural diversity of the pathogen. One key difference between our work and these is that unlike the other methods our approach discovers the potentially meaningful structure of the sequences.

From the perspective of this paper, one of the key aspects of the acquired immune response of the mammalian immune system is its ability to distinguish short ($<\sim 25$) amino acid sequences in pathogen proteins (non-self) from those found in host proteins (self). In this context then, the dependencies among nearby sequence positions captured by our approach are more important than longer range dependencies.

## 2.2 SNP ANALYSIS

A *single nucleotide polymorphism* (SNP) is a variation in a DNA sequence that involves a single nucleotide. That is, it is a position in the DNA sequence of a species where different members may have different nucleotides. SNPs account for a significant part of the human genetic variation with a SNP occurring on average about every 600 nucleotides. There is an keen interest in discovering associations between SNPs and phenotype, such as disease and drug response.

It has been observed that SNPs group into segments of limited diversity (Patil et al., 2001). Within one of these *haplotype blocks* there are far fewer multi-SNP patterns than would be expected were the SNPs in the block were independent. These are exactly the kinds of dependencies captured by the hidden variables discovered by our method. Many of the key SNP/haplotype block computational problems correspond to structural properties or simple inferences in our learned models:

- Locating the boundaries of haplotype blocks (Anderson & Novembre, 2003; Zhang et al., 2002). Given SNPs from different individuals, the task is to identify the haplotype block structure. In our models the block boundaries are given by the segmentation. Interestingly one of these approaches (Anderson & Novembre, 2003) use the Minimum Description Length principle to identify the block structure. Our experiments find MDL to be the least effective of the three scoring functions we consider. The approach of Zhang et al. (2002), like ours, uses a dynamic program. Their approach, however, does not entail learning a probabilistic model of SNP sequences.

- Predicting missing SNPs (Su et al., 2005). The process of determining an individual's SNPs is noisy and frequently has missing data. Additionaly, current technologies only have the density to identify a small fraction of all known SNPs. Prediction of missing SNPs in our models is done by performing inference, a simple and fast procedure given the structure of our models.

- Finding a set of representative SNPs (Chang et al., 2006). This task involves finding a small set of SNPs (called "Tag SNPs") whose values determine (with some confidence) the values of all other SNPs in some set. In our models this problem involves finding a set of observed variables in a segment that have a high information gain about that segment's hidden variable.

Perhaps the closest work to ours in the SNP field is that of Greenspan & Geiger (2004). Their approach also involves learning probabilistic models of sequences. The main strutural difference is that their models contain directed edges between neighboring hidden variables while ours do not. Additionally, their approach to learning model structure involves greedy search while ours is based on finding the optimal segmentation.

## 3 MODELS AND METHODS

Each candidate hidden variable we consider is the parent of a contiguous *segment* of observed variables. For example, $H_1, H_3$ and $H_6$ in Figure 1(b). We identify a candidate HV by specifying both its segment and its number of states or cardinality. The model search space consists of all partitions of the observed variables into a set of non-overlapping segments along with cardinalities.

We call members of this search space *segmentations* where it is understood that there must be accompanying cardinalities. The segmentation

$$S = ((s_1, l_1, c_1), ..., (s_a, l_a, c_a), ..., (s_M, l_M, c_M))$$

contains $M$ segments where $s_a$ is the start position of segment $a$, $l_a$ is its length and, if $l_a > 1$, $c_a > 1$ is the cardinality of $H_a$, the hidden variable associated with segment $a$. If $l_a = 1$ then $X_a$ has no parents (e.g., $X_3$ in Figure 1(b)) and $c_a$ is ignored. We require the segmentation to be complete and non-overlapping, so $s_1 = 1$, $S_M + l_M = L + 1$ and $s_a + l_a = s_{a+1}$. We also use $e_a = s_a + l_a - 1$ to refer to the end position of segment $a$.

Given a segmentation, we assume that the sequences in each segment are marginally independent so,

$$\Pr(\mathbf{X}|S) = \prod_{a=1}^{M} \Pr(\mathbf{X}_a)$$

where $\mathbf{X} = \{X_1, ..., X_L\}$ are the observed variables and $\mathbf{X}_a = \{X_{s_a}, ..., X_{e_a}\}$ are the observed variables in segment $a$.

We model sequences within a segment longer than 1, which we call a correlated segment, with a multinomial mixture model

$$\Pr(\mathbf{X}_a) = \sum_{k=1}^{c_a} \Pr(H_a = h^k) \times \prod_{i=s_a}^{e_a} \Pr(X_i | H_a = h^k)$$

where the $h^k$ are the states of $H_a$ and the conditionals $\Pr(X_i | H_a = h^k)$ are unrestricted multinomials. For length 1 segments we have $\Pr(\mathbf{X}_a) = \Pr(X_{s_a})$ where $\Pr(X_{s_a})$ is an unrestricted multinomial.

The objective of our hidden variable discovery method is to discover the "best" segmentation according to a suitable segmentation scoring function.

### 3.1 FINDING THE OPTIMAL SEGMENTATION

Due to the independence of segments, the score of a segmentation for each of the three scoring methods we consider (see below) decomposes into a sum of scores of the individual segments:

$$\mathsf{score}(S, \mathbf{x}) = \sum_a \mathsf{seg\_score}(s_a, l_a, c_a, \mathbf{x}_a)$$

where $\mathbf{x}$ is a training set of sequences and $\mathbf{x_a}$ is the part of $\mathbf{x}$ aligned with segment $a$.

Below we address the issue of scoring a segment. Now, we show how to efficiently find the optimal segmentation.

The optimal segmentation, $S^*$, for a training set is the one that maximizes $\mathsf{score}$:

$$S^* = \underset{S \in \mathcal{S}}{\operatorname{argmax}} \, \mathsf{score}(S, \mathbf{x})$$

where $\mathcal{S}$ represents the set of all segmentations.

Since, ignoring cardinalities, there are $2^{L-1}$ possible segmentations exhaustive search for even moderate length sequences is prohibitive. Because $\mathsf{score}$ decomposes, however, there is an efficient dynamic program (DP) for finding $S^*$.

The DP fills two length $L$ vectors, $\vec{V}$ and $\vec{W}$, from 1 to $L$. The value of $V(i)$ is the score of the maximum scoring segmentation of the first $i$ positions $(X_1, ..., X_i)$ and $W(i)$ holds the length and cardinality of the last segment in the optimal segmentation of $(X_1, ..., X_i)$. We begin by setting $V(0) = 0$ and then fill the rest of $\vec{V}$ and $\vec{W}$ for $i = 1$ to $i = L$ according to the recursion

$$V(i) = \max_{(l,c)} V(i-l) + \mathsf{seg\_score}(i-l+1, l, c, \mathbf{x})$$

$$W(i) = \underset{(l,c)}{\operatorname{argmax}} V(i-l) + \mathsf{seg\_score}(i-l+1, l, c, \mathbf{x}).$$

After $V$ and $W$ have been filled, the optimal segmentation is constructed starting with the last segment given by $W(L)$ and tracing back to the first segment. Given the segment scores, the computational complexity of the DP is $O(L^2 C)$ where $C$ is the maximum cardinality allowed. The DP can be easily modified if, as in our experiments, there is a maximum allowed segment length.

### 3.2 SCORING A SEGMENT

Scoring a segment is the problem of scoring a mixture model given the cardinality of its hidden variable. In related work (Cheeseman & Stutz, 1996; Chickering & Heckerman, 1997) such a scoring function is used (in part) to set the cardinality of the hidden variable of a mixture model. For these tasks, a scoring function is good if it gives the highest score to a good cardinality, however, the properties we desire in a segment scoring function are a bit different because we want a scoring function whose scores combines well (via the DP) to yield a good overall segmentation.

We consider three segment scoring functions: the *Bayesian Information Criterion* (Schwarz, 1978) (or alternatively the *Minimum Description Length* (Rissanen, 1983)) (BIC/MDL) score, the *Cheeseman-Stutz* (CS) (Cheeseman & Stutz, 1996) score and the *cross-validation* (CV) score.

Both the BIC/MDL and CS scores are based on large sample estimates of the *marginal likelihood*. A common score used to guide BN structure search is the log of the joint posterior probability of the structure and training data, $\log \Pr(S, \mathbf{x}) = \log \Pr(S) + \log \Pr(\mathbf{x}|S)$. Assuming uniform structure priors, the $\Pr(\mathbf{x}|S)$ term, called the marginal likelihood, is used to score model structures[1]. Since computation of this term requires integration over all model parameter settings and is often intractable if there are hidden variables, approximations of the marginal likelihood are commonly used.

The BIC score of segment $S_a = (s_a, l_a, c_a)$ is

$$\mathsf{seg\_score}_{\mathsf{BIC}}(S_a, \mathbf{x_a}) = \log \Pr(\mathbf{x}_a | \hat{\boldsymbol{\Theta}}) - \frac{d \times \log N}{2}.$$

Here $N$ is the number of training set sequences. The first term is the likelihood of the training data at $\hat{\Theta}$, the maximum likelihood estimate of the parameters of the model, and the second term is a complexity penalty that contains $d$, the dimension of the model equal to the number of free parameters, $d = (c - 1) + ((A - 1) \times c \times l)$ where $A$ is the alphabet size. One problem with BIC is that, since $d$ scales with the alphabet size, if the alphabet has infrequently observed symbols or, equivalently, if most of the mass in the multinomial distributions concentrates on few symbols, the penalty term can be too severe. The BIC approximation is equal to minus the Minimum Description Length criterion.

---

[1] We consider the cardinalities of hidden variables part of the model structure

The CS score of segment $S_a$ is

$$\text{seg\_score}_{\text{CS}}(S_a, \mathbf{x_a}) =$$

$$\log \Pr(\mathbf{x_a}, \mathbf{h'_a}) + \log \Pr(\mathbf{x_a}, \mathbf{h'_a}|\tilde{\Theta}) - \log \Pr(\mathbf{x_a}|\tilde{\Theta})$$

where $h_a$ are the "settings" for $H_a$ for each sequence in $\mathbf{x_a}$ so that the MAP estimate $\tilde{\Theta}$ for the completed data set $(\mathbf{x_a}, \mathbf{h_a})$ is equal to the MAP estimate for $\mathbf{x_a}$. The term $\Pr(\mathbf{x_a}, \mathbf{h'_a})$ is the marginal likelihood of the complete data, and can be computed efficiently (Heckerman, 1995) under certain assumptions, one of which (parameter independence) clearly does not apply to at least one of our domains.

Unlike the BIC/MDL and CS scores, the CV score is not based on an approximation of the marginal likelihood, but rather on an empirical estimate of generalization. To compute the CV score we partition the training set into $k$ disjoint sets and estimate MAP parameters of $k$ models, withholding one of the $k$ sets from the training set each time. The CV score is the log likelihood of held aside sequences

$$\text{seg\_score}_{\text{CV}}(S_a, \mathbf{x_a}) = \sum_{j=1}^{N} \log \Pr(\mathbf{x_a^j}|\tilde{\Theta}_j)$$

where $\mathbf{x_a^j}$ is segment $a$ in sequence $\mathbf{x^j}$ and $\tilde{\Theta}_j$ are the MAP parameters learned when $\mathbf{x_a^j}$ is withheld.

The necessity of learning the parameters of $k$ models can be expensive with general BN structures. With the simple mixture models we consider, however, inference, and thus learning, is reasonably efficient. So, while the computational cost of CV is more than both BIC/MDL and CS, it is not prohibitive.

To estimate the maximum likelihood or MAP parameters, we use the *expectation-maximization* (Dempster et al., 1977) or EM algorithm. Since EM converges to a local maximum, we use random restarts (10 in our case) retaining the model with the highest ML or MAP score.

## 3.3 PRACTICAL MATTERS

Although each of the segment scoring functions are rather efficient, there are $O(L^2 C)$ candidate segments, and for longer sequences the cost of scoring every segment can be irksome, at the least. There are a number of techniques that can reduce the time to do a segmentation. Most importantly, perhaps, the segment scores can be calculated in parallel. Given realistic resources, however, there is a need to score at least some potion of the segments serially. We have used two techniques to speed these calculations First, for any subsequence, its score as a function of the cardinality should[2] have a single local maximum, thus if the score drops considerably with an increase in cardinality, scoring segments for that subsequence can cease. Second, segment scores should never decrease if the segment is extended by one, $(\text{seg\_score}(s, l, c) \leq \text{seg\_score}(s, l+1, c)$ We can use this relation to avoid scoring segments that cannot be in the optimal segmentation.

## 4 Empirical Evaluation

We evaluate our approach on two real world data sets. The first data set, which we call vaccine, contains aligned amino acid sequences of VAR2CSA, an antigenic protein expressed by the parasite that causes malaria (*Plasmodium falciparum*), assembled by our collaborators in an ongoing malaria vaccine project (Bockhorst et al., 2007). The vaccine data set has 106 length 2859 sequences from a 21 symbol alphabet[3], 23% of the sequence matrix is observed, the rest is missing data. The second data set, SNP, contains SNP data from human chromosome 21 (Patil et al., 2001). We evaluate our approach on a 260 position long stretch studied previously (Su et al., 2005). This data set contains 20 length 260 sequences from a two letter alphabet {1,2} where 2 denotes the majority SNP and 1 denotes the minority SNP. 683 (13%) entries of this data set are missing, and of the observed data 27% of the SNPs are in the minority class.

All iterations of EM are run with 10 random restarts except when learning the final mixture models following segmentation when we use 25 random restarts. We use a Dirichlet prior with parameters $1.0/A$ (the alphabet size) for all emission parameters in the observed variables and no pseudo counts on the parameters of hidden variables. On vaccine we set the maximum cardinality to 10 and the maximum segment length to 15. On SNP the maximum cardinality is 5 and the maximum segment length is 50.

---

[2] "should" because of the stochastic nature of EM with random restarts

[3] 20 amino acids plus a special "gap" symbol that results from the alignment process.

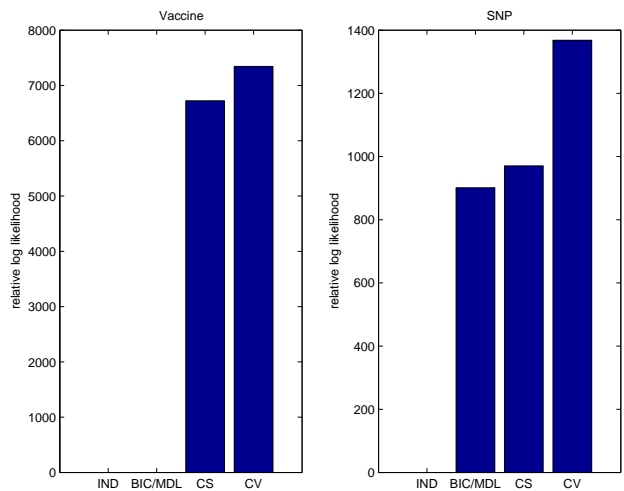
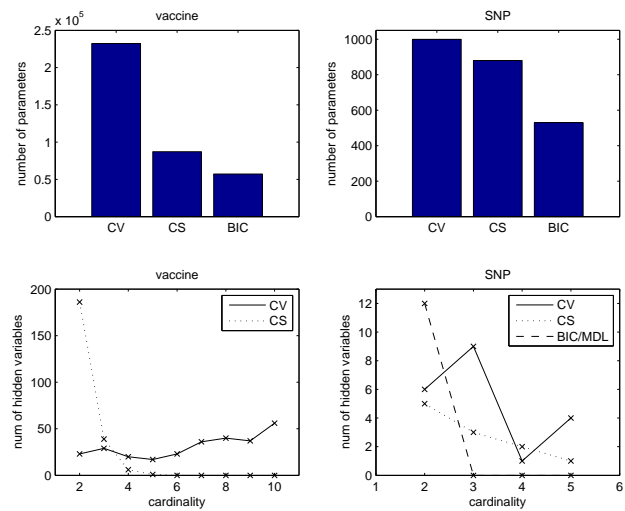

Figure 2: Test set log likelihood relative to the baseline (IND) which assumes all observable variables are independent.

Figure 3: Properties of models learned from BIC/MDL, CS and CV scores on vaccine and SNP tasks using all training data. Shown above are plots of the number of free-parameters. Shown below are histograms for the cardinalities of the hidden variables.

We have conducted a set of experiments in order to answer several key questions: i) How well do models learned with our approach predict unseen sequences? ii) How do models trained with the three segment scoring functions, BIC/MDL, CS and CV, compare? iii) What is the value the dynamic program for recovering the optimal segmentation? iv) How do models trained with our approach compare to the state-of-the art on the task of predicting missing SNPs?

To address the first two questions, we compare learned models in a 10-fold cross-validation experiment[4]. We use our DP to construct maximum scoring segmentation models using segment scores based on each of the three segment scoring functions. As a baseline we train a model that assumes that the observed variables are independent. Figure 2 shows the test set log-likelihood for the three segment scoring functions using the baseline as a reference. On both tasks the CV scoring method results has the highest log likelihood.

CV assigned a higher likelihood than CS to 84 of 106 vaccine sequences and 17 of 20 SNP sequences and CV assigned a higher likelihood than BIC/MDL to 103 of 106 vaccine sequences and 16 of 20 SNP sequences ($p$-value from binomial test all $< 0.01$). We conclude that the CV score yields more accurate models than either BIC/MDL or CS on these tasks. While the CS based models outperform the BIC/MDL models on both tasks (consistent with previous results on individual mixture models (Chickering & Heckerman, 1997)), it is the poor performance of BIC/MDL on vac-cine which is most striking. In fact the models learned with BIC/MDL on this task have no hidden variables. and thus BIC/MDL is identical to the baseline.

---

[4]Note that here cross-validation refers to withholding test set sequences to evaluate our model.

The difference in behavior of MDL/BIC on the two tasks is likely due to the difference in alphabet sizes. While an increase in cardinality by one to a HV for a length $l$ segment adds $l$ free parameters in SNP, it adds $20 \times l$ parameters in vaccine. Consequently, a candidate segment in vaccine incurs a much greater complexity penalty than a similar segment in SNP. Since many positions in vaccine have only one or two kinds of observed letters, and thus have a similar complexity to SNP, a position specific complexity penalty may be needed for BIC/MDL to be competitive on this task. Figure 3 shows the number of parameters and cardinality histograms for models learned with all the training data. Not surprisingly BIC/MDL models have the fewest hidden variables with high cardinalities (and fewer parameters) and CV models have the most.

To assess the value of the models learned with the DP approach we compare them to models learned using a greedy approach. For both approaches we use CV based segment scores. The greedy approach first computes a normalized segment score by dividing the CV segment score by the segment length. Next it constructs a segmentation by repeatedly adding the non-overlapping segment with the highest normalized score until all positions are covered. In addition to the independent baseline, we consider a cluster baseline (CLUST). This naive Bayes model has a single hidden variable which is a parent of all observed variables. We set the cardinality ($\geq 2$) of the hidden variable in a CLUST model using the CV method.

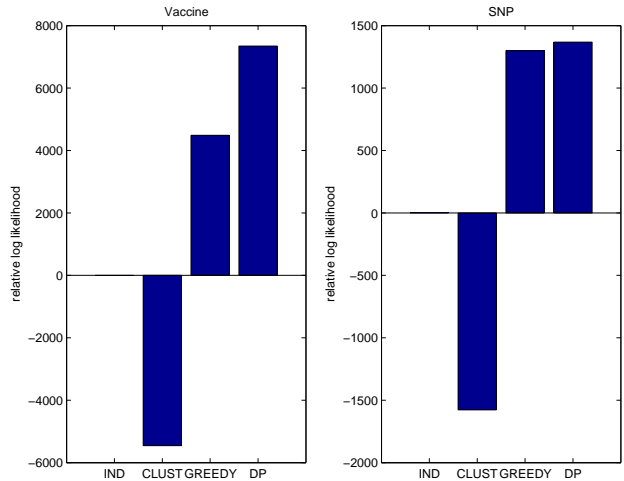
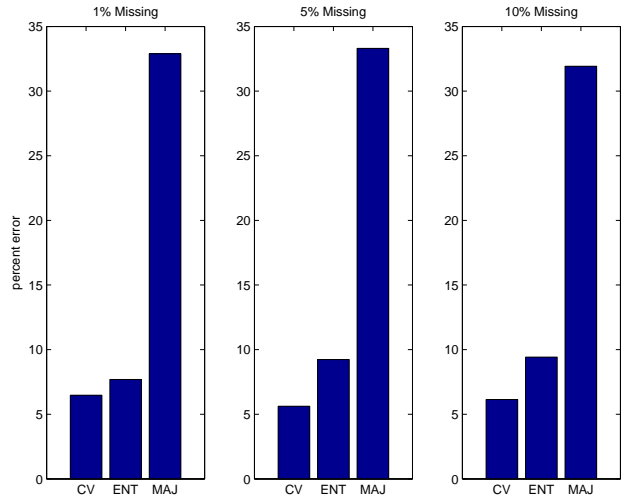

Figure 4: Test set log-likelihoods relative the baseline model (IND).

Figure 5: Error rates for the task of predicting missing SNPs.

Figure 4 shows the test set log likelihoods for these approaches. On both tasks models based on the optimal segmentation obtained by DP has the highest test set log likelihood. On the vaccine task DP assigns a higher log likelihood than GREEDY on 96 of 106 sequences ($p$-value $< 0.001$) while on the SNP task DP assigns a higher log likelihood than GREEDY on just 12 of 20 sequences. Thus the DP approach outperforms the GREEDY method on vaccine, and the results are mixed on SNP. Interestingly, vaccine sequences are much longer than SNP sequences, and there are thus more places for the myopic GREEDY algorithm to make sub-optimal choices. The CLUST method has the worst performance, even compared to IND, indicating that there is no strong relationship among the sequence (in either task) that holds across the whole length. An assumption of this kind of relationship is implicitly made by most techniques for building phylogenetic trees of biological sequences.

The last task we consider is that of predicting values for missing SNPs. For missing rates of 1%, 5% and 10% we trained segmentation models from data sets in which randomly chosen SNPs were hidden. We used CV segment scoring along with the DP approach. We then predicted the value of the missing SNPs by performing inference with the learned models. This was repeated 10 times for each missing rate. Figure 5 shows the mean error rates of our approach along with the reported error rates of a state-of-the-art approach on this same data (Su et al., 2005). We label this approach ENTROPY because it is based on minimizing the within segment entropy. We also show the error rate of an approach that predicts the majority SNP. The standard deviations of the error rate using our approach are 2.2%, 1.0% and 1.3% for 1%, 5% and 10% missing rates. Our CV/DP approach has the lowest error rate for all three missing rates. Unlike ENTROPY, its error rate does not change significantly as the percent missing increases from 1% to 10%. The difference between the error rate of our approach and ENTROPY increases as the % missing grows, and with 10% of SNPs missing the error rate of CV/DP is over 35% less than the state-of-the art ENTROPY method. There are two key differences between our method and ENTROPY that may account for this improvement. First, ENTROPY requires a complete data set for training, which they acquire by filling in missing values with the majority SNP value, while our approach deals naturally with missing data. Second, ENTROPY does not allow for subtle differences among sequences with the same type. If, for example, two sequences from a long segment differ in only one position, ENTROPY treats those sequences as different types and would consider the distance between them as great as the distance from each to a third sequence with completely different SNP values. Our approach, on the other hand, has a flexible notion of a type as given by the mixture model.

## 5 Conclusion

We have described a simple approach to discovering patterns of local correlations in sequential data sets. Our approach uses a dynamic program to find the optimal way to partition the observed variables into a set of non-overlapping and independent mixture models.

We evaluated our approach on two computational biology domains. One set of sequences came from a vaccine design domain and the other from the analysis of SNPs. Our empirical evaluation shows:

- Models constructed from the optimal segmentation generalize better than models learned with a greedy approach, especially on the vaccine task which contains longer sequences.

- Models built from the cross-validation score had better test set performance that models built from either BIC/MDL or Cheeseman-Stutz scores. The BIC/MDL models had the poorest performance. BIC/MDL appears to have been hindered by an overly severe model complexity penalty.

- On the task of predicting missing SNP values, our approach reduced the prediction error rate of a state-of-the-art method by up to 35%.

One intriguing research direction we are exploring is to use a segmentation to provide an alternate representation of biological sequences as a series of types and mutations. This representation naturally leads to alternate methods to calculate distances between sequences, which become especially important in recombining sequence families. For example, we can base the distance measure on only type differences, only differences within the same type or a combination of both. We are actively exploring how these techniques may be used to uncover relationships among populations of rapidly evolving and recombining sequence families.

### Acknowledgements

This research was supported in part by the University of Wisconsin - Milwaukee Graduate School. Thanks to Craig Struble for suggesting applications of our methods to SNP data.